\documentclass[10pt]{article} 
\usepackage[accepted]{tmlr}

\usepackage{amsmath,amsfonts,bm}

\def\eqref#1{equation~\ref{#1}}

\def\1{\bm{1}}

\DeclareMathAlphabet{\mathsfit}{\encodingdefault}{\sfdefault}{m}{sl}
\SetMathAlphabet{\mathsfit}{bold}{\encodingdefault}{\sfdefault}{bx}{n}

\usepackage{hyperref}
\usepackage{url}

\usepackage{soul}
\usepackage{amsthm}
\usepackage{booktabs}

\usepackage[linesnumbered,vlined,ruled]{algorithm2e}
\usepackage{listings}
\usepackage[most]{tcolorbox}
\usepackage{multicol}
\usepackage[normalem]{ulem}

\newtheorem{definition}{Definition}
\usepackage{stmaryrd} 
\usepackage{subcaption} 
\usepackage{cleveref}
\usepackage{makecell}

\newcommand{\enc}[1]{\llbracket #1 \rrbracket}

\let\cite\citep

\title{Labeling without Seeing?\\Blind Annotation for Privacy-Preserving Entity Resolution}

\author{\name Yixiang Yao \email yixiangy@usc.edu \\
      \addr Department of Computer Science\\
      University of Southern California
      \AND
      \name Weizhao Jin \email weizhaoj@usc.edu \\
      \addr  Department of Computer Science\\
      University of Southern California
      \AND
      \name Srivatsan Ravi \email srivatsr@usc.edu\\
      \addr  Department of Computer Science\\
      University of Southern California}

\def\month{05}  
\def\year{2025} 
\def\openreview{\url{https://openreview.net/forum?id=bAM8y3Hm0p}} 

\begin{document}

\maketitle

\begin{abstract}
The entity resolution problem requires finding pairs across datasets that belong to different owners but refer to the same entity in the real world. To train and evaluate solutions (either rule-based or machine-learning-based) to the entity resolution problem, generating a ground truth dataset with entity pairs or clusters is needed. However, such a data annotation process involves humans as domain oracles to review the plaintext data for all candidate record pairs from different parties, which inevitably infringes the privacy of data owners, especially in privacy-sensitive cases like medical records. To the best of our knowledge, there is no prior work on privacy-preserving ground truth labeling in the context of entity resolution. We propose a novel blind annotation protocol based on homomorphic encryption that allows domain oracles to collaboratively label ground truth without sharing data in plaintext with other parties. In addition, we design a domain-specific, user-friendly language that conceals the complex underlying homomorphic encryption circuits, making it more accessible and easier for users to adopt this technique. The empirical experiments indicate the feasibility of our privacy-preserving protocol (f-measure on average achieves more than 90\% compared with the real ground truth).
\end{abstract}

\section{Introduction}
\label{sec:intro}

\begin{figure}[!hbt]
    \centering
    \includegraphics[width=0.5\linewidth]{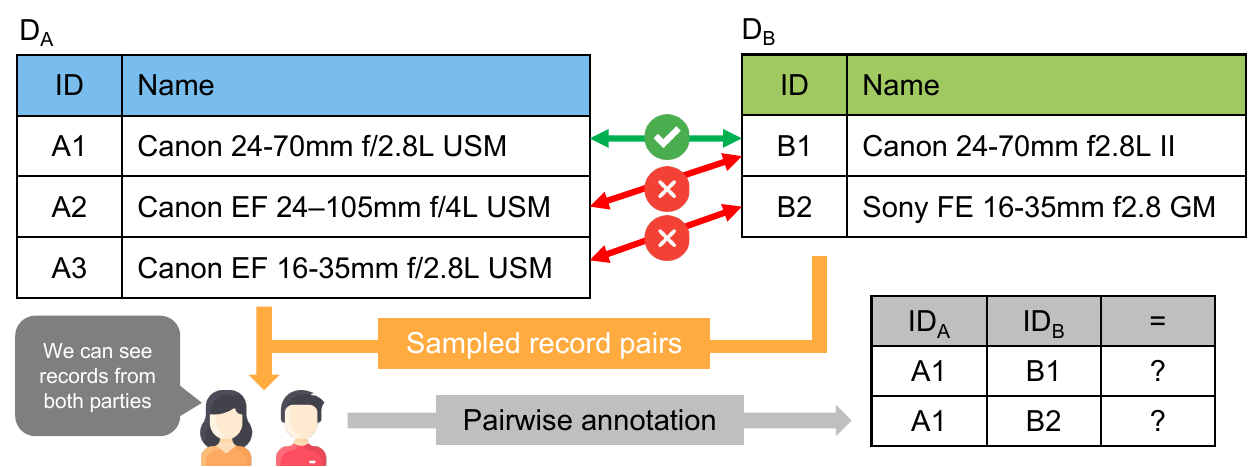}
    \caption{The entity resolution and annotation example. $D_A$ and $D_B$ are two lens product datasets owned by two parties separately. Making pairwise annotation requires the \textcolor{orange}{visibility} of both parties' sampled records, revealing private data.}
    \label{fig:er-example}
\end{figure}

\textit{Entity resolution} is the task of linking entity pairs across datasets that refer to the same entity in the real world. It is an essential problem in massive data integration and quality improvement, which has been widely employed in various domains including academia, industries, and government agencies~\cite{christen2011survey,winkler2014matching}. Traditionally, it requires the participation of multiple owners of the data sources, and it is unavoidable that the data is shared among parties. As the example shows in \Cref{fig:er-example}, datasets $D_A$ and $D_B$, which belong to two different parties, consist of the IDs and names of the digital camera lens products but in different representations. Each lens is an entity and the goal is to link the same entities between these two datasets. To achieve that, the data owners of $D_A$ and $D_B$ share the data with each other and jointly determine that \texttt{A1} from $D_A$ and \texttt{B1} from $D_B$ could be a match since they are all ``Canon 24-70mm f2.8'' camera lenses though \texttt{A1} explicitly notes the lens is with ``USM'' motor and \texttt{B2} is the "II" generation. On the contrary, \texttt{A2} and \texttt{B1} are not a match since they have different focal ranges (``24-105'' vs. ``24-70'') and aperture sizes (``4'' vs. ``2.8''), as well as \texttt{A3} and \texttt{B2} since they belong to different brands (``Canon'' vs. ``Sony''). 
As real-world entity data usually contains sensitive personal information or is related to intellectual property, \textit{privacy-preserving entity resolution} (PPER)~\cite{gkoulalas2021modern,yao2021amppere,ghai2023lessons,yao2024feasibilityprivacypreservingentityresolution}, as a solution, is to link entities with the protection for data privacy. The linking process does not reveal any unnecessary information and the parties will not learn any entities that are not common among all parties. Back to our example of lens products, $D_A$'s owner only knows \texttt{A1} has a potential match \texttt{B1} in $D_B$ but does not learn anything about \texttt{B2}. Likewise, $D_B$'s owner also learns no other lens products from $D_A$ except the match with \texttt{A1}. 

In the lifecycle of entity resolution and PPER, a ground truth dataset with entity pairs or clusters needs to be prepared to evaluate the quality of linking or train learning-based entity resolution models. Constructing such a ground truth dataset, known as \textit{annotation}, requires the involvement of domain oracles (data owners or trusted delegators)~\footnote{In practice, domain oracles are individuals who possess a deep understanding of datasets, enabling them to make high-quality annotations.}. Normally, the first step is to sample some representative entities from all data sources and convert them into pair-wised forms. The oracles then manually determine if such a candidate ``pair'' is considered to be legitimate in the real world. For example, in \Cref{fig:er-example}, assume that all records are sampled, oracles could recognize \texttt{(A1,B1)} is a match but \texttt{(A1,B2)} is not. Annotation tools~\cite{ipeirotis2010demographics,neves2014survey} including Amazon Mechanical Turk~\cite{buhrmester2016amazon,kang2014privacy,lease2013mechanical} are for constructing and curating ground truth datasets. 

However, this is a human-in-the-loop process and the oracles need to see all the raw data of entities side-by-side as pairs in plain.
It is worth pointing out that even though most of the privacy-aware entity resolution algorithms themselves are considered to be somewhat privacy-preserving, the ground truth labels they used, no matter what strategies are conducted for minimizing the size of the required labels, are still annotated in plaintext~\cite{kurniawan2022homomorphic} and can not be identified as \textit{privacy-preserving annotation}. For example, healthcare facilities aim to perform entity resolution on their confidential medical records, yet the process of creating accurate reference annotations still relies on conventional annotation tools that lack robust privacy guarantees.

Recently, several privacy-preserving annotation methods have been proposed. While some methods~\cite{kajino2014instance,zhang2021privacy} employ perturbation for crowd-sourced annotations, others leverage existing AI-driven ground truth generation in specific tasks~\cite{kavasidis2012semi,wangt2001automatic}, they are limited to certain domains~\cite{d2009semi,jing2017automatic} and almost implausible to have such techniques universally generalized for privacy-preserving entity resolution. Additionally, the differential privacy-based methods~\cite{feyisetan2019} protect the Personal Identifiable Information (PII) of individual records, but the content remains in plaintext.
Contrary to the relatively well-researched privacy-preserving entity resolution solutions for the application phase~\cite{schnell2009privacy,scannapieco2007privacy,inan2008hybrid,inan2010private}, there still exists a privacy protection void in the annotation phase for PPER. Making annotations for entity resolution privacy-preserving is challenging because the domain oracles must compare the data across parties \textit{side-by-side} in order to determine if two records are the same. Moreover, the raw data can certainly be manipulated accordingly or the annotation process can be semi-auto for the purpose of privacy protection, but the \textit{partial content} of candidate pairs is still inevitably revealed for oracles at least need to see some of the plaintext to understand the meaning and then make decisions.

\begin{figure*}[!hbt]
    \centering
    \includegraphics[width=.9\linewidth]{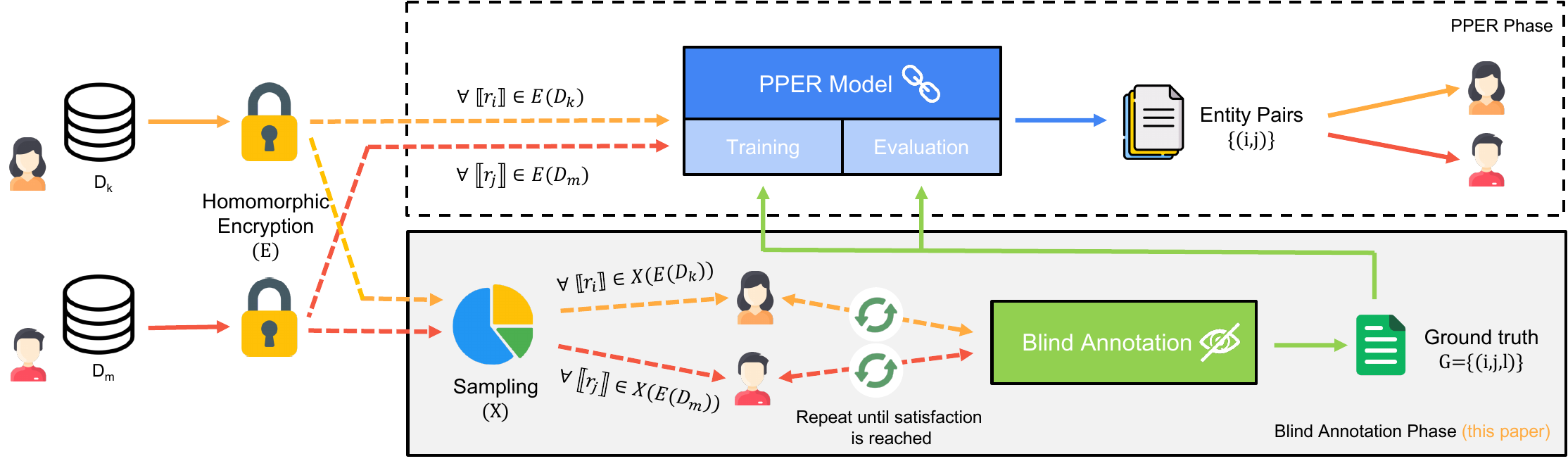}
    \caption{The relation between privacy-preserving entity resolution (upper \dashuline{white dashed box}) and blind annotation (lower \underline{grey solid box}). In this paper, we focus on the annotation phase and utilize ``blind annotation'' to label ground truth for PPER with \textit{zero-knowledge} shared with other parties. The homomorphically encrypted records from each party are first sampled. Then, the data owners or their delegated oracles annotate sampled records \textit{without} looking at the other party's data but interacting with the blind annotation protocol. Finally, the ground truth as a set of triplets $G=\{(i,j,l)\}$ ($i$ and $j$ are record ids from two parties and $l$ is a label indicating if $r_i$ and $r_j$ is a match) is formed and can be used for training and evaluation in any PPER task. Note that no record content $r_i$ or $r_j$ in plaintext is revealed throughout the annotation process. The \underline{solid line} denotes the data stream in plaintext and the \dashuline{dashed line} denotes the data stream in ciphertext. $D$ denotes dataset,  $X$ denotes sampling, $E$ denotes homomorphic encryption, and $\enc{r}$ denotes record $r$ in ciphertext.}
    \label{fig:conceptual-model}
\end{figure*}

Different from the aforementioned approaches that are not suitable for pairwise privacy annotation or merely hide the individual identifier but not the content, we propose a novel ``blind annotation'' approach of ground truth labeling, based on homomorphic encryption (HE)~\cite{fontaine2007survey,acar2018survey}, for privacy-preserving entity resolution where oracles from each party can only inspect the data of their own but are still able to work collaboratively. The conceptual model is as shown in \Cref{fig:conceptual-model}, where data owners "blindly" annotate the ground truth set, which can later be used in the model training and evaluation phases of privacy-preserving entity resolution. No data in the plain view of each party is shared with any other parties so the privacy of the data is guaranteed without loss of data utility.

It is important to clarify that this paper does \textbf{NOT} aim to address PPER itself. Instead, we propose a novel \emph{privacy-preserving method for pairwise ground truth annotation}, which is applicable to any PPER pipeline. The relationship between PPER and the proposed privacy-preserving annotation method is illustrated in \Cref{fig:conceptual-model}.
We summarize our main contributions as follows: 

\begin{itemize}
    \item We propose a novel homomorphic-encryption-based privacy-preserving protocol that allows multiple parties to collaboratively produce pairwise entity annotations without sharing their data with any other parties in plaintext. To the best of our knowledge, this is the first solution to discover the potential of \emph{privacy-preserving ground truth annotation for pairwise entity annotations} in entity resolution.
    \item The proposed protocol allows the domain oracles to focus solely on the annotation logic, and the protocol
handles the optimizations for the underlying homomorphic circuit. Each party involved in the protocol can perform the computation efficiently and independently. The privacy properties of the protocol have been rigorously proven.
    \item We implement the protocol and conduct empirical studies on heterogeneous real-world datasets to prove the feasibility of our blind annotation protocol.
\end{itemize}

\section{Problem Definition}

The \textbf{privacy-preserving entity resolution} (PPER) problem can be defined as a triple $T=(D, M, E)$ where $D = \{D_1 \ldots D_n\}$ is a set of $n$ different datasets consisting of records $r$, from $n$ different data owners $P = \{P_1 \ldots P_n\}$ respectively. $E$ denotes encoding or encryption algorithm that keeps the $r$ from each $D$ in encoded or ciphertext form, that is, $\enc{r} \in E(D)$, where $\enc{r}$ explicitly denotes $r$ is in ciphertext. $M$ is the match set containing record pairs between any two datasets amongst the $n$ parties, so 
$M=\{(\enc{r_i}, \enc{r_j})\,|\,r_i = r_j; \enc{r_i} \in E(D_k), \enc{r_j} \in E(D_m)\}$,
where $r_i=r_j$ indicates that $r_i$ and $r_j$ refer to the same entity in the real world, despite having different representations.



Finding the complete and precise $M$ depends on the domain-specific algorithm/model which relies on the high-quality ground truth data for training, fine-tuning or evaluation. The ground truth set $G$ is a set of triples $(i, j, l)$ where $i$ and $j$ are record ids of a record pair $r_i$ and $r_j$, and $l$ is a Boolean label which indicates if such record pair is the same entity. The process of constructing $G$ is called \textbf{annotation}. Formally, 
$G=\{(i, j, l)\,|\,r_i \in X(D_k), r_j \in X(D_m)\}$,
where $X$ is a sampling algorithm. $l$ is determined by domain oracles with the content of $r_i$ and $r_j$. 

Usually, $G$ is not constructed individually from each party: because the domain oracles have to see $r_i$ and $r_j$ in clear in order to determine whether two records are the same real-world entity based on the features the records have. Therefore, the owners of the records need to share sampled records in the plain with other parties which makes the raw content of $r_i$ and $r_j$ revealed. 

To prevent the potential privacy leakage during this process, this paper focuses on creating ground truth data with $\enc{r_i}$ and $\enc{r_j}$ straight so that no plaintext from any parties reveals to any other parties. We name this process \textbf{blind annotation}, that is,
$G=\{(i, j, l)\,|\,\enc{r_i} \in X(E(D_k)), \enc{r_j} \in X(E(D_m))\}$.

\section{Preliminaries}
\subsection{Related Works}


\noindent\textbf{Privacy-preserving annotation} As stated in this privacy-preserving record linkage survey paper\cite{gkoulalas2021modern}, almost no open method is available for making ground truth annotation for PPER due to privacy concerns regardless of its a time-consuming and erroneous process.

The only similar scenario we found that explored the attempts for privacy-preserving annotation is about active learning using differential privacy~\cite{feyisetan2019}. The general active learning method aims to learn a distribution of the data by selecting less training data that carries the highest information with the active learner. The sampled data from the learner needs to be annotated by domain oracles or crowd workers. The problem is that crowdsourcing the labels is usually from an open call and transmitting non-public data to crowd workers has an inevitable privacy leakage risk, especially for the identities. For the insufficient privacy guarantee of $k$-anonymity, the authors adopted a differential privacy algorithm~\cite{li2012sampling}, which prevents a user from sustaining additional damage by including their data in a dataset, on binary classification tasks, this method achieves similar accuracy scores as the non-privacy counterpart with a small performance hit but strong privacy guarantee. Though Personal Identifiable Information (PII) is protected from revealing, the content of the sampled data as plain text is still known by the crowd workers for them to understand and annotate. If the content of the non-public data itself requires to be kept private, such a method is not yet qualified.

Another method to prepare data for machine learning tasks is to learn the distribution of the original data and generate synthetic data based on that~\cite{el2020seven}. Even though this method might work for training statistical models and conducting analytical works, it is not feasible to be applied to entity resolution tasks. Because the representation of the token/word is an important signal to determine the similarities between records, a tiny modification or substitution of the original representation could cause huge judgment deviations for the annotators.

\citet{yu2020hyper} provided a method of not considering ground truth but instead using unsupervised heuristic measures based on a greedy matching approach to evaluate and optimize the hyper-parameters for entity resolution models. This method is based on the assumption that a match of record pairs gets the highest similarity score among a set of heuristic measures against other candidate pairs. Using heuristics to estimate linkage quality is doable in some certain scenarios, however, evaluating heuristics by heuristics, in general, is somewhat a ``the chicken or the egg'' problem. Additionally, this method only works in some general cases where the same entity looks similar: in some extreme conditions or some hard record linking problems, the representations of the record are different even when they are referring to the same entity.

In short, labeling ground truth in a privacy-preserving manner is tough and no direct work exists. Our approach achieves the goal without compromising privacy protection by employing homomorphic encryption.

\noindent\textbf{Comparison to PSI and PIR}
Private Set Intersection (PSI)~\cite{morales2023private} or weighted PSI are general-purpose solutions for identifying common elements between sets. However, they have limited extensibility, particularly when it comes to supporting complex annotation logic. Specifically, PSI does not natively support conditional expressions or complex predicates, and incorporating such logic typically requires substantial additional effort. In contrast, the blind annotation protocol can express more sophisticated logical operations more naturally.

Moreover, PSI and weighted PSI can still be used within blind annotation protocol for computing set intersections. Although they are not currently available in our DSL, they can be incorporated as functions, provided that the PSI operation can be represented by HE circuits~\cite{chen2017fast}.

Private Information Retrieval (PIR)~\cite{chor1998private}, on the other hand, is designed for retrieving data without disclosing which item is being accessed. While state-of-the-art PIR extensions support more expressive queries, such as keyword-based retrieval~\cite{cryptoeprint:2025/210}, they share the same limitation as PSI in that they do not easily accommodate complex logical operations. In comparison, the blind annotation protocol provides a more flexible foundation for such logic.

\subsection{Homomorphic Encryption}
\textit{Homomorphic encryption} (HE) allows the computation to be performed over encrypted data while preserving the input/output relationship of the function between the plaintext and ciphertext data~\cite{fontaine2007survey,acar2018survey}. In general, an encryption scheme is said to be homomorphic if for some operator $\odot$ over plaintext ($\odot_{\mathcal{M}}$) and ciphertext ($\odot_{\mathcal{C}}$), the encryption function $E$ satisfies: $\forall m_1, m_2 \in \mathcal{M},\ E(m_1 \odot_{\mathcal{M}} m_2) \leftarrow E(m_1) \odot_{\mathcal{C}} E(m_2)$, where $\mathcal{M}$ denotes the message in plaintext and $\mathcal{C}$ denotes the message in ciphertext. $\leftarrow$ declares the computation is direct without any decryption in the middle of the process. Therefore, if we let $\odot$ to be $+$, computing $m_1+m_2$ can be done by a computation unit that receives the original message from the data owner in the encrypted form, that is, $E(m_1)$ and $E(m_2)$, and is able to compute the addition of two encrypted messages without decrypting. Only the data owner with the key can decrypt the message from ciphertext back to plaintext.


Figure~\ref{fig:he-operations} demonstrates a typical public key homomorphic encryption scheme $\varepsilon$ primarily characterized by four operations. \textit{Key generation} takes in a security parameter $\lambda$ and outputs public key pair $(pk, sk)$. \textit{Encryption} encrypts plaintext message $m$ with $pk$ and returns ciphertext $c$. \textit{Decryption}, as an inverse, decrypts ciphertext $c$ with $sk$ into plaintext $m$. \textit{Evaluation} evaluates a function (represented as a circuit) $f$ over a ciphertext tuple $(c_1, c_2, \dots ,c_k)$ with $pk$ and returns $c_f$, which is equivalent to $f(m_1,m_2,\dots,m_k)$ after decrypting.

\begin{figure}
\setlength\tabcolsep{0pt} 

\begin{tcolorbox}[colframe=black!70!white,colback=black!10!white]
\renewcommand{\arraystretch}{1.2} 
\begin{tabular}{l l}
\textbf{\texttt{Key Generation}:\ } & $(pk, sk) \leftarrow \texttt{HE.KeyGen}_{\varepsilon}(\lambda)$ \\
\textbf{\texttt{Encryption}:} & $c \leftarrow \texttt{HE.Enc}_{\varepsilon}(m, pk)$ \\
\textbf{\texttt{Decryption}:} & $m \leftarrow \texttt{HE.Dec}_{\varepsilon}(c, sk)$ \\
\textbf{\texttt{Evaluation}:} & $c_f \leftarrow \texttt{HE.Eval}_{\varepsilon}(f, (c_1, c_2, \dots ,c_k), pk)$ \ \ s.t. $\texttt{HE.Dec}_{\varepsilon}(c_f, sk) \rightarrow f(m_1,m_2,\dots,m_k)$ \\
\end{tabular}
\end{tcolorbox}
\caption{Operations of a typical public-key homomorphic encryption scheme $\varepsilon$}
\label{fig:he-operations}
\end{figure}

The party that executes the homomorphic functions does not know anything about the data and the party that provides the encrypted data is unaware of what functions have been executed. This property, which is informally named \textit{blind evaluation}, is the basis of our protocol.
\section{Blind Annotation Protocol}
\label{sec:protocol}

We propose a privacy-preserving annotation protocol that allows the ground truth to be annotated among multiple parties ``blindly'' without inspecting other parties' records.

\noindent\textbf{Intuition:} The intuition behind this is that without putting a candidate record pair side-by-side for domain oracles to determine if they are the same entity or not, the oracles on each side could extract the core features by just looking at the record content in plaintext of their own and summarize these features into a series of Boolean questions. If any record from the other parties satisfies all conditions defined by these questions, this record is highly likely to be a match of the record that the questions are derived from. Suppose the record content is encrypted, and these Boolean functions can be blindly evaluated over the ciphertext of the record. In that case, no plaintext of the record content is revealed to other parties.

For example, from $P_A$'s perspective, determining if a $P_B$'s record \texttt{B1} is similar to its record \texttt{A1} can be based on \texttt{A1}'s core features. If \texttt{B1} also has keywords ``Canon'', ``24-70'' and ``2.8'' as \texttt{A1}, it is highly-likely that it is a potential match to \texttt{A1}. The other keywords like ``mm'', ``USM'' are not necessary for identifying the lens based on the oracle's domain knowledge. Figuring out these features only needs $P_A$ access to its own record \texttt{A1}. Moreover, if $P_A$ encapsulates the feature detection functions for \texttt{A1} to be homomorphic functions and $P_B$ encrypts \texttt{B1}, $P_A$ can label record pair \texttt{(A1,B1)} using homomorphic evaluation without knowing the content of \texttt{B1}.

\begin{figure}[!hbt]
    \centering
    \includegraphics[width=.9\linewidth]{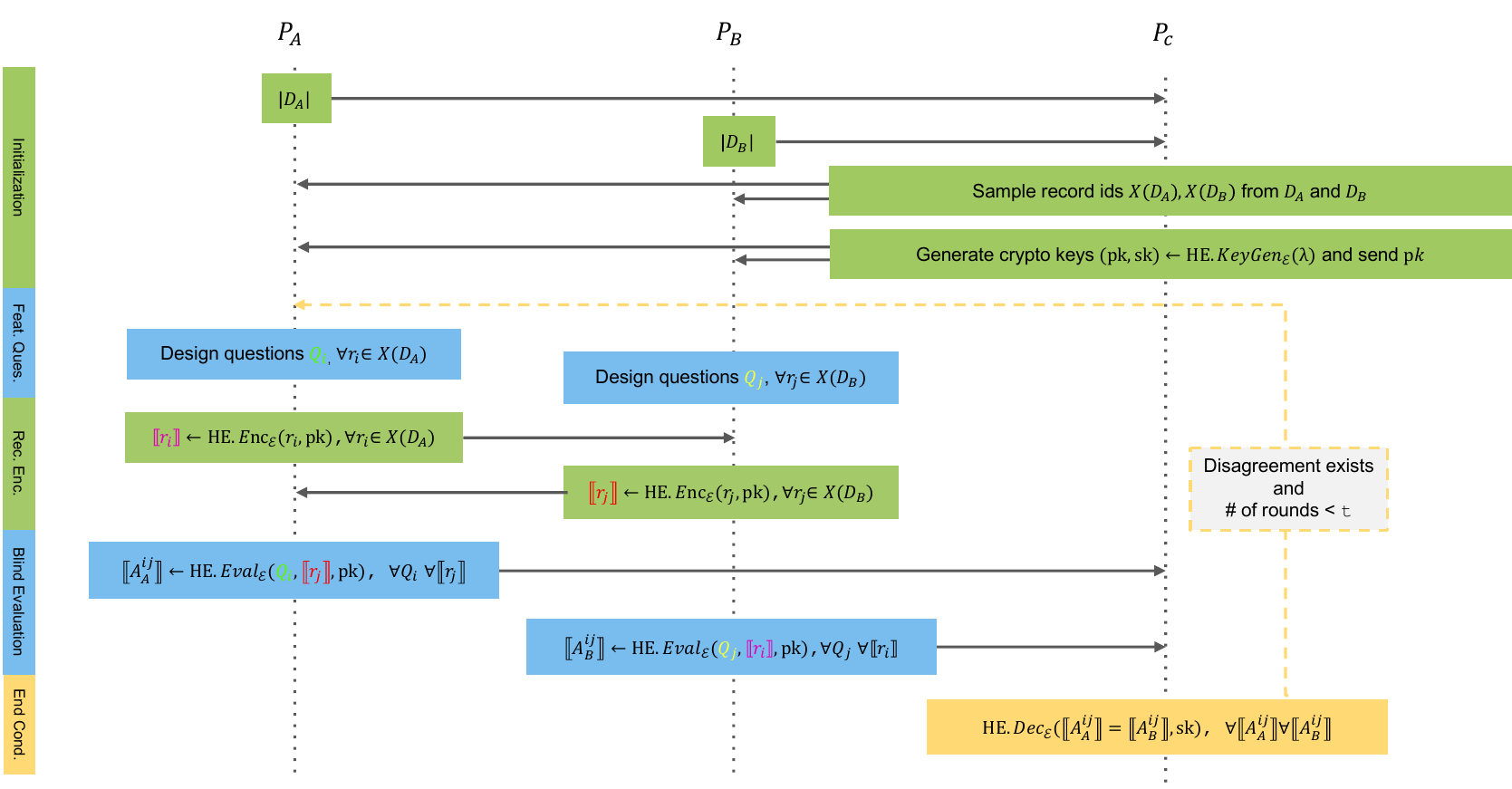}
    \caption{The timing diagram for the blind annotation protocol. The protocol, as well as five main components (initialization, feature questions, record encryption, blind evaluation, and end conditions) are dissected in \Cref{sec:protocol}.}
    \label{fig:timing}
\end{figure}

\noindent\textbf{Protocol:} The protocol is elaborated in \Cref{fig:timing}. Succinctly, assume the ground truth construction is between party $P_A$ and $P_B$, and party $P_C$ is a coordinator for key management and result collection. In the \textit{initialization} phase, party $P_A$ and $P_B$ first send their dataset size to $P_C$, and $P_C$ randomly samples the record ids. Moreover, $P_C$ also prepares the public key pair $(pk, sk)$ for homomorphic encryption and sends the public key $pk$ to $P_A$ and $P_B$, and keeps the secret $sk$ for decryption.

The next step is called \textit{feature questions}, which is for data owners or domain oracles to annotate their records using homomorphically executable functions. Concretely, $P_A$ and $P_B$ sample records $r_i$ and $r_j$ respectively according to the sampled record ids from $P_C$, and prepare a set of Boolean logic-style questions $Q_i$ and $Q_j$ accordingly to the features of the record content for each record. Each question set $Q$ is combined into a form that returns one Boolean result with first-order logic and converted to a homomorphically executable function.

Following that is \textit{record encryption}, where the clear record data is homomorphically encrypted with $pk$ in $P_A$ and $P_B$ respectively, and the ciphertext $\enc{r_i}$ and $\enc{r_j}$ are shared with another party.

The core step is \textit{blind evaluation}, where the encrypted records from the other party are evaluated with prepared feature questions. Concretely, Once $P_A$ receives a $\enc{r_j}$ from $P_B$, it evaluates $Q_i$ over it homomorphically, that is $\texttt{HE.Eval}_{\varepsilon}(Q_i, \enc{r_j}, pk)$, and gets $\enc{A_A^{ij}}$. $P_B$ does a similar operation and gets $\enc{A_B^{ij}}$. Both $\enc{A_A^{ij}}$ and $\enc{A_B^{ij}}$ are in ciphertext and sent to $P_C$. $P_C$ then tests if $\enc{A_A^{ij}}$ equals $\enc{A_B^{ij}}$ and decrypts the result with the secret key $sk$.

The last step is to determine the \textit{end conditions}. If $P_A$ and $P_B$ agree on the result, the result will be stored. Otherwise, this pair will be picked out for the next round. In the next round, $P_A$ and $P_B$ require to refine their questions and evaluate with the encrypted record data from another party again. After $t$ rounds, the pairs with no agreement will be discarded.

\noindent\textbf{Privacy:} From $P_A$'s perspective, throughout the protocol, $P_A$ has no access to $P_B$'s record content $r_j$ in plaintext but evaluates questions over ciphertext $\enc{r_j}$. The question functions $Q_i$ are evaluated on $P_A$'s side so $P_B$ does not know anything about the features tested in $Q_i$ by $P_A$. The same situation applies to $P_B$. $P_C$, which has the secret key $sk$, decrypts the final results which contain no record content in plaintext from either $P_A$ or $P_B$. None of the parties is able to apply ciphertext collision attack because homomorphic encryption is semantically secure (\Cref{sec:record-encryption}).

In the following subsections, we walk through the protocol details and dissect each essential component. We also provide the protocol security analysis in \Cref{sec:security}.
Note that parties operate independently throughout the process as long as the interactions follow the timeline in the \Cref{fig:timing}. 
Although Secure Multi-Party Computation (MPC) is a potential option for implementing the blind annotation protocol, \emph{HE is preferred in practice due to its ability to support offline evaluation}. This is especially important because each party may perform annotation at different times, with varying time costs, and HE does not require all parties to remain online concurrently.

\subsection{Initialization}

Prior to getting involved in the annotation, a few initialization steps should be conducted first. The dataset owners ($P_A$ and $P_B$) need to first consent on the \textbf{annotation criteria} including how similar can two records be identified as the match and what kind of difference is tolerable. For example, consider the records ``Canon 24-70mm f/2.8L USM'' and ``Canon 24-70mm f2.8L USM II'' (\texttt{A1} and \texttt{B1} in \Cref{fig:er-example}). Should these two lenses, which differ in generation, be treated as the same? In some contexts, lenses with generation I and II should be distinguished, while in others they may be categorized as the same item. The data used for deciding such criteria in this step can be synthetic or be derived from the record content so it does not leak any (sensitive) information about the original records.

A \textbf{data preprocessing} step is necessary for some of the datasets. If the dataset owners believe the format of their data has a noticeable distinction, they could apply a series of data cleaning and standardization operations, for example, removing the dataset-specific characters, lemmatization or stemming for natural language content, on their data individually.

$P_C$ is responsible for \textbf{sampling records} from $P_A$ and $P_B$'s dataset. After the dataset size $|D_A|$ and $|D_B|$ are sent to $P_C$, $P_C$ randomly samples some amount of record from $D_A$ and $D_B$ and sends back the selected record ids as a list back to the original data owners respectively. $P_C$ then \textbf{generates the key pairs} $(pk, sk) \leftarrow \texttt{HE.KeyGen}_{\varepsilon}(\lambda)$ with selected homomorphic encryption scheme $\varepsilon$ and parameters $\lambda$, then distributes the public key $pk$ to both $P_A$ and $P_B$ for encrypting their data and questions. Note that $P_C$ is the \textit{only} party that can decrypt any ciphertext back to plaintext with access to the secret key $sk$.

Note that all steps, except for HE key generation, are also required in a standard \emph{non-private annotation setup} and are essential for the final annotation quality. Our proposed method preserves these steps unchanged.Only after the annotation criteria have been defined and consistent data preprocessing has been applied can appropriate feature questions be formulated for the records.

\subsection{Feature Questions}

Feature questions are designed to test whether essential features of a record are satisfied and are annotated by domain oracles. This is based on the assumption that if all the core features of two records are the same, then they are highly likely to be the same entity. Specifically, the features of the records could be the \emph{sub-strings} that are necessary for representing the record. For example, in a lens name ``Canon 24-70mm f/2.8L USM II'', one feature can be the focal length ``24-70'' which can be used to construct a question that tests if ``24-70'' is in the compared lens. Therefore, using this question to evaluate a ``15-85'' lens returns false.

Under this assumption, the data owner can design a set of Boolean questions for all the features of a certain record. Utilizing first-order logic chains a set of questions together and the evaluation of a record turns it into a single Boolean value as the final result. For example, assume the question set is $Q=\{Q_1, Q_2, \ldots Q_n\}$, and it could be $Q=Q_1 \vee \neg Q_2 \ldots \wedge Q_n$ after conversion with first-order logic. Therefore, $Q(r)=Q_1(r)\vee \neg Q_2(r) \ldots \wedge Q_n(r)$ returns a Boolean value that implies whether the record $r$ has all the desired features.

In our protocol, the feature question set $Q$ is as $f$ in homomorphic encryption that uses to evaluate the encrypted record $\enc{r}$ with \texttt{HE.Eval}, that is, $\texttt{HE.Eval}_{\varepsilon}(Q, \enc{r}, pk)$, or simply $Q(\enc{r})$. Constructing $Q$ that is homomorphically executable is \emph{non-trivial} due to the operator limitation and the hardness of the efficient encryption circuit construction (see details in \Cref{sec:functions}). 

Hence, on a high level, we design a simple annotation language for the ease of use in \Cref{sec:dsl} and pre-defined some primitive functions that encapsulate the details of underlying homomorphic circuits in \Cref{sec:functions}. With this layered design, \emph{the domain oracles focus solely on the annotation logic}, and the program handles the tricks and optimizations for the sophisticated underlying homomorphic circuit construction.

\subsubsection{Domain-specific Language}
\label{sec:dsl}

The annotation is designed to be written in a domain-specific language (DSL). This language provides general methods for data manipulation and logic computation, which are extensible for defining functions. 
The syntax of the DSL is in \Cref{sec:dsl-syntax}.

With this DSL, it is efficient and sufficient to "ask questions". Take the lens example, the record content from $P_A$ is ``Canon 24-70mm f/2.8L USM II'', the annotation from the oracle could be:

\begin{lstlisting}
$r = lower($r)
$c1 = is_in("canon", $r)                         # condition 1
$c2 = is_in("24-70", $r) | is_in("2470", $r)     # condition 2
$c3 = !is_in("24-105", $r)                       # condition 3
ret $c1 & $c2 & $c3
\end{lstlisting}

where target record \texttt{\$r} is first being lower-cased, and then requires the brand to be ``Canon'', focal range to be ``24-70'' or simply ``2470'' as the common abbreviation but can not be ``24-105''. Therefore, both ``Canon 24-70 f2.8'' and ``Canon 2470'' are considered matches, but ``Canon 24-105mm USM'' isn't. Notice not all features but the important ones, that are consensus in the annotation criteria, are tested, e.g., the motor type ``USM'' or the version ``II'' are not being used for making the decision.

\subsubsection{Homomorphic Functions}
\label{sec:functions}

Writing homomorphic functions from scratch is not as straightforward as writing some normally "obvious" functions because of all the cumbersomeness and restrictions that homomorphic encryption schemes bring:
(1) The homomorphic encryption scheme only supports a collection of limited operators. More complicated functions need to be built using these basic building blocks.
(2) Simply porting the logic of a normal function to a homomorphic encryption function would suffer from privacy and/or efficiency issues. The function needs to revamp or refactor for it to work properly.
(3) To achieve desired privacy protection (\Cref{sec:security}), the encryption is end-to-end, so no decryption is allowed in the middle of the evaluation, which confines the use of control flow in programming.


To address these challenges and for the ease of writing effective and efficient functions along with the proper privacy protection, we pre-define several functions that are essential components for constructing Boolean questions $Q$ and can be evaluated homomorphically without exposing the details of the underlying circuits. For the encryption schemes that only support logical operators, the arithmetic operators could be enriched by building upon the lower-level gate circuits, for example, constructing an 8-bit adder with AND/OR/NOT/XOR gates~\cite{mano2017digital}. On the other hand, the arithmetic schemes such as BGV can also be extended to support logical comparisons~\cite{iliashenko2021faster} (e.g., ``<'' and ``='') and that also retrains the benefit of efficient SIMD operations~\cite{smart2014fully} naturally come with these schemes.

It is worth mentioning that, among all the predefined functions, one function is \texttt{is\_in($\enc{a}$, $\enc{b}$)} that tests if the encrypted string $\enc{a}$ is a sub-string of the encrypted string $\enc{b}$. This is the core to measure if a certain feature $\enc{a}$ exists in a record $\enc{b}$.
In terms of extensibility, as long as a function needed can be encapsulated in a way that is able to compute homomorphically, it is safe to be added to the framework.

\subsection{Record Encryption}
\label{sec:record-encryption}

To make the blind evaluation functions work, the records from both data owners require to be encoded with the same data encoding method. For the current case, since we only work on English characters, we simply apply ASCII encoding over the original records. Each character in a record is then represented as an 8-bit integer. All the sampled records $r \in X(D)$ are encrypted with the public key $pk$ that $P_C$ generated in the initialization step. 

The output ciphertext $\enc{r} \leftarrow \texttt{HE.Enc}_{\varepsilon}(r, pk)$ is different even if the content of $r$ is exactly the same due to the probabilistic encryption property, also known as semantic security, that homomorphic encryption carries~\cite{yi2014homomorphic}. For instance, encrypting ``Canon 24-70mm'' $n$ times results in $n$ different ciphertexts. Therefore, no party is even able to test the exact match of the records by comparing the encrypted record ciphertext from other parties. 

Finally, the encrypted $\enc{r} \in X(E(D))$ is sent to another data owner for blind evaluation.

\subsection{Blind Evaluation}

From $P_A$'s perspective (the perspective for $P_B$ is interchangeable), once it receives the encrypted $X(E(D_B))$ from $P_B$, $\enc{r_j} \in X(E(D_B))$ is fed into each prepared function $Q_i$ for blind evaluation. Computing $Q_i$ with $\enc{r_j}$ is a typical homomorphic evaluation process that $P_A$ does not know anything about what the content is in $\enc{r_j}$ but can still evaluate it as the plaintext version $r_j$ with all the questions defined in $Q_i$. Because $P_A$ does not know which $\enc{r_j}$ potentially matches its own record $r_i \in X(D_A)$, this process needs to test all $\enc{r_j}$s against all $Q_i$s. Therefore, the total number of such process executes $|D_A| \times |D_B|$ times for $|Q_i|=|D_A|$.

The output $\enc{A^{ij}_A} \leftarrow \texttt{HE.Eval}_{\varepsilon}(Q_i, \enc{r_j}, pk)$ for each $\enc{r_j}$ is also in ciphertext and since the secret key $sk$ is only accessible by $P_C$, $P_A$ is not possible to get any information from $X(E(D_B))$. Note $\enc{A^{ij}_A}$ associates with the record id pairs $(i,j)$ so that $P_C$ is able to identify the provenance of the result.

\subsection{End Conditions}

The encrypted results from $P_A$ and $P_B$ are collected and evaluated by $P_C$. On $P_C$'s side, it aligns $\enc{A^{ij}_A}$ and $\enc{A^{ij}_B}$ according to $i$ and $j$, and tests if $\enc{A^{ij}_A}=\enc{A^{ij}_B}$ meanwhile decrypts the result with secret key $sk$. $P_C$ then has a mapping $F$ with record id pairs $(i,j)$ as keys and Boolean values indicating if two parties have made an agreement on the same record pairs as values. Formally,
$F=\{(i,j) \mapsto \texttt{HE.Dec}_{\varepsilon}(\enc{A^{ij}_A}=\enc{A^{ij}_B}, sk)|\ r_i \in X(D_A), r_j \in X(D_B)\}$.

For each record pair $(i,j)$ in $F$, if the value is true, the agreement has been established between the two data owners, so no additional process is needed. Otherwise, the record pair needs further investigation, and $P_C$ extracts all $i$s and $j$s from disagreed pairs and returns them to the data owner respectively. The data owners have to conduct another round of annotation only for these records. The annotations for the later rounds tend to be not as strict as the former rounds so it increases the possibility of making $\enc{A^{ij}_A}$ and $\enc{A^{ij}_B}$ the same. Simultaneously, $P_C$ maintains another list $G_h$ that stores the annotation result from one of the parties (assuming $P_A$ here), that is,
$G_h=\{(i,j,l)\ |\ h \in [1, t], l \leftarrow \texttt{HE.Dec}_{\varepsilon}(\enc{A^{ij}_A}, sk)\}$,
where $l$ is the label, $h$ is the round number and $t$ is a parameter denoting the maximum number of rounds that should be decided between the data owners in the initialization step. When $t$ rounds have finished, the protocol ends: the pairs whose value is true (consensus archived) in $F$ are added to the final ground truth, and others are discarded. Therefore, the ground truth set $G$ is constructed as
$G = \{(i,j,l)\ |\ (i,j,l) \in G_t, F(i,j)=true\}$.
Note that increasing $t$ tends to improve performance, but it also raises the labeling cost. An empirical analysis of the effect of $t$ is presented in \Cref{sec:exp-feasibility}.

Obviously, when all the values in $F$ are true in the $h$-th round, it is by no means to recurse into the $(h+1)$-th round. The protocol meets the early termination condition and exits immediately.

\section{Experiments}

In this section, we conduct experiments to empirically evaluate the feasibility of using blind annotation to annotate datasets and the incurred overhead of homomorphic encryption.
In general, domain oracles are asked to annotate datasets using blind annotation. The quality of the final annotation results is assessed by comparing them to the ground truth.

\begin{table*}[ht]
\centering
\small
\renewcommand{\arraystretch}{1.1} 
\begin{tabular}{rcl|cc|cc}
\toprule
&&&\multicolumn{2}{c|}{Original} & \multicolumn{2}{c}{Selected} \\
Dataset & Domain & Attributes & \#E & \#M & \#E & \#M \\
\midrule
Abt-Buy & E-commerce & \underline{name}, description, manufacturer, price & 1081+1092 & 1097 & 50+50 & 50 \\
Amazon-Google & E-commerce & \underline{name}, description, manufacturer, price & 1363+3226 & 1300 & 49+50 & 50 \\
DBLP-Scholar & Bibliographic & \underline{title}, \underline{authors}, \underline{venue}, \underline{year} & 2616+64263 & 5347 & 49+50 & 50 \\
DBLP-ACM & Bibliographic & \underline{title}, \underline{authors}, \underline{venue}, \underline{year} & 2614+2294 & 2224 & 50+50 & 50 \\
\midrule
Febrl & Biomedical & \makecell[l]{\underline{given name}, \underline{surname}, \underline{sex}, \underline{age}, \underline{title}, \\ date of birth, \underline{address 1}, address 2, \\ \underline{phone}, soc sec id, culture, family role} & 1500+3500 & 1074& 50+50 & 50 \\
\bottomrule
\end{tabular}
\caption{Dataset specifications and basic statistics. The \underline{underlined attributes} are the attributes selected in the experiments. \#E represents the number of entities, and \#M denotes the number of matches.}
\label{tab:dataset}
\end{table*}

\subsection{Settings, Datasets and Metrics}


\subsubsection{Settings.} To empirically study the feasibility of our blind annotation protocol in a practical sense, we implement a web-based user-friendly GUI and its corresponding HE program (see details in \Cref{sec:simulator}). Specifically, the web GUI is implemented in Python, in which the DSL syntax is written in Extended Backus–Naur Form (EBNF) and parsed by Lark library~\footnote{https://github.com/lark-parser/lark} with Look-Ahead Left-to-Right (LALR) parser. The syntax definition can easily be extended for new syntax, operators, and functions. The workflow is as follows: each domain oracle~\footnote{Given that these tasks require only general knowledge, the oracles are individuals who understand the datasets.} annotates the owned dataset individually, and the annotation program merges and calculates the results. If the annotations do not satisfy the exit condition, the unqualified records are returned to the corresponding oracles for the next annotation.

The underlying HE program is implemented in OpenFHE~\cite{al2022openfhe}, an open-source project that efficiently and extensibly implements the post-quantum Fully Homomorphic Encryption schemes. We use the \texttt{BinFHE} module, a concrete implementation of FHEW/TFHE~\cite{10.1007/978-3-662-46800-5_24,chillotti2020tfhe}. Specifically, we set it to work in public-key encryption mode with the crypto context to be \texttt{STD128}, which guarantees more than 128 bits of security for classical computer attacks. Additionally, we implement the program in two versions: serial and parallel. The latter employs OpenMP~\footnote{https://www.openmp.org/}, a multi-platform shared-memory parallel programming library, for accelerating independent gate operations to be executed in parallel.
All the experiments are conducted on a Linux machine with an 8-core CPU @ 3.60 GHz and 32 GB RAM.

\subsubsection{Datasets.} We use the real-world entity resolution benchmark~\cite{kopcke2010evaluation}, which includes 4 tasks and lies in both e-commerce and bibliographic domains. Each task consists of two datasets and a ground-truth file, which contains all the true matches. To mimic the labeling process, we extract a subset of records from each dataset, annotate them, and then compare the annotated pairs to the ground truth. Specifically, we first randomly sample 50 labeled matches from each provided ground truth, and this covers at most 5\% (50 records) of each dataset because one record could link to multiple records. Note that 50 records from each dataset are around 2.5-5\% of the original dataset except for Scholar. Even though the proportion of sampled records as ground truth seems small, it is \textbf{sufficient} for ER in most practical scenarios~\cite{kasai2019low}. 
The specifications and the basic statistics of the dataset are listed in \Cref{tab:dataset}.
We do not make any specific data cleaning and normalization except that all the Unicode characters are mapped into the ASCII range.

Generally, PPER are evaluated using ER's benchmark datasets. However, an important application area is in healthcare, particularly with patient data. Unfortunately, these datasets often require permission for use or are restricted to specific purposes, making them less ideal for evaluating the effectiveness of the privacy-preserving methods. As an alternative, synthetic datasets are commonly used. One such synthetic dataset is Febrl (Freely Extensible Biomedical Record Linkage) \cite{christen2008febrl}, which is widely employed for generating census records containing fields such as name, sex, age, and address. These fields align with the structure of many patient datasets. Furthermore, as a synthetic dataset, it allows for customizable levels of noise. For our experiments, we generated the experiment dataset using the settings from \citet{yao2021amppere}.
The record content contains typos, ORC errors and phonetic festures, and we summarize the statistics of missing value in \Cref{tab:febrl-missing-value}.

\begin{table}[ht]
\centering
\small
\begin{tabular}{ccccccc}
\toprule
given name & surname & sex & age & title & address 1 & phone \\
\midrule
1.8 & 1.3 & 21.2 & 19.6 & 57.5 & 5.7 & 5.6 \\
\bottomrule
\end{tabular}
\caption{The percentage (\%) of missing value for each attributes in the Febrl dataset}
\label{tab:febrl-missing-value}
\end{table}

\subsubsection{Evaluation Metrics.} We use precision, recall and F-measure as the standard evaluation metrics to measure the accuracy of blind annotation against the scores computed from ground truth labels. We additionally analyze other observed phenomenons, including the relationship between the annotations rounds and the number of labeled pairs that come to an agreement. As for the homomorphic implementation, the communication and computation costs are evaluated.

\begin{figure*}[!hbt]
    \centering
    \includegraphics[width=.99\linewidth]{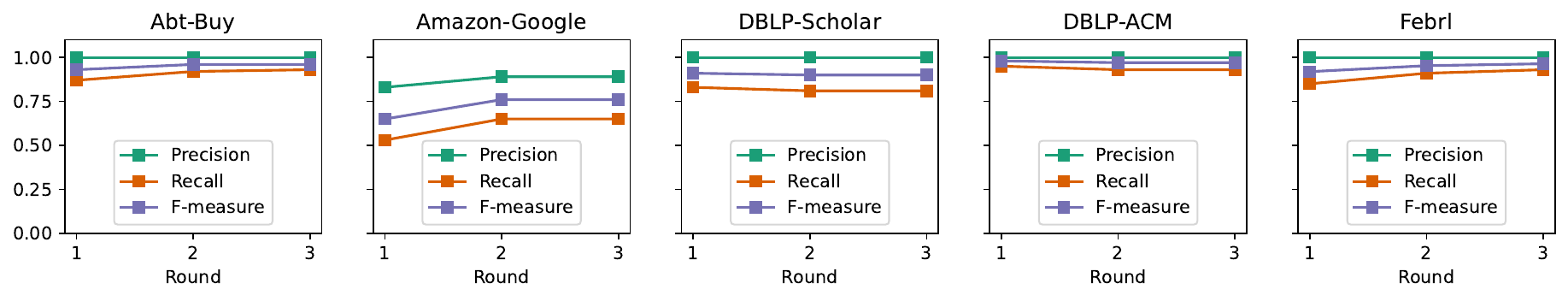}
    \caption{Performance evaluation regarding precision, recall and f-measure.}
    \label{fig:exp-performance}
\end{figure*}

\begin{figure*}[!hbt]
    \centering
    \includegraphics[width=.99\linewidth]{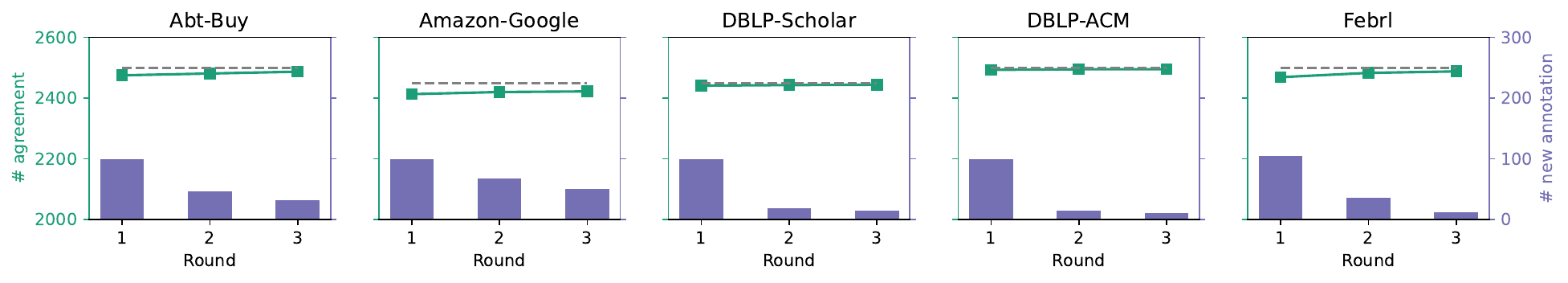}
    \caption{\#round to meet agreement. The gray dashed line is the total number of candidate pairs, and the green solid line is the actual number of pairs that meet agreement. These two use the left y-axis as the scale. The purple bars at the bottom are the summation of the number of records to annotate from two parties. It uses the right y-axis as the scale.}
    \label{fig:exp-agreement}
\end{figure*}

\subsection{Feasibility Verification}
\label{sec:exp-feasibility}

We try to validate two hypotheses in the following experiments to verify the feasibility of using blind annotation to construct the ground truth dataset for PPER: 
\begin{itemize}
    \item (H1) The feature extraction from only one side of the dataset without inspecting the other side and comparing candidate pair side-by-side is a feasible approach.
    \item (H2) More annotation rounds are effective for improving the quality of the PPER ground truth labeling.
    \item (H3) It is efficient for domain oracles to learn the DSL and perform annotation through the GUI.
\end{itemize}



\noindent\textbf{H1} From \Cref{fig:exp-performance}, we observe the F-measure for Abt-Buy, DBLP-Scholar, and DBLP-ACM are all above 0.9, but Amazon-Google is not as good as others. The Amazon-Google dataset used in our experiment can be considered as hard problems for the record representations are full of abbreviations (e.g., software as sw), missing brands or models for products, and a limited amount of information (very short names). If better data pre-processing, especially normalization, is applied beforehand, a visible performance boost should be achieved. With more rounds of records annotated, precision and f-measure increase sharply for Abt-Buy and Amazon-Google, but not that noticeable for DBLP-Scholar and DBLP-ACM since both of them have already achieved high scores at the initial stage. Surprisingly, the recall values for DBLP-Scholar and DBLP-ACM drop slightly with more rounds executed, this is mainly due to the side-effect of the annotation strategy: when annotations from two parties do not come to an agreement, the oracles tend to make the annotations to be more generalized, that is, relaxing on the strictness of the matching criterion for easier capturing similarities, in the next round. Normally, having slack criterion will diminish precision but rise recall for more negatives become positives; however, another variable, the number of agreements between parties, is also increasing in blind annotation, which leads to the increment of both true positives and false negatives and the recall somewhat drops as a consequence. For Febrl, performance has a boost. It finally achieves promising results, largely due to the presence of highly identifiable signals, such as names, phone numbers, and keywords in addresses, in the census data.

\noindent\textbf{H2} We further explore other effects that occur with the iteration of annotation rounds. As is shown in \Cref{fig:exp-agreement}, the green line, which indicates the number of agreements, approaches the gray line, which is the total number of candidate pairs, as the annotation round increases. On the other hand, the number of records that requires to annotate, shown as the bars, decreases. Specifically, for Abt-Buy and Amazon-Google, the annotation demands drop to 1/4 and 1/2 of the original amount respectively; for DBLP-Scholar and DBLP-ACM, in the second and the third round, this number is even less than 10\%, because of the more explicit features the datasets have. For Febrl, the agreement increases notably across rounds, primarily due to the progressive relaxation of the criteria related to missing values.


\noindent\textbf{H3} The domain oracles became familiar with the DSL and GUI in around 30 minutes. For labeling, the average time per record was under 1 minute using the web-based GUI (implementation details are provided in \Cref{sec:simulator}). The annotation process was fully offline for each party and executed in parallel.

In conclusion, the first hypothesis holds because all evaluation results lie in the acceptable range. With better cleaned and normalized datasets, the performance in terms of precision, recall and f-measure should be promising. The second hypothesis holds because precision, recall, and f-measure surge at the beginning with the increment of annotation rounds, and tend to be steady after a series of such rounds. The number of disagreements between the parties drops over the rounds before it hits the plateau. The third hypothesis holds as well, as both the learning and annotation costs remain within an acceptable range.

\subsection{HE Overheads Analyses}
\label{sec:he-overheads}

\begin{figure}
\centering
\begin{subfigure}[t]{.48\textwidth}
  \centering
  \includegraphics[width=1\linewidth]{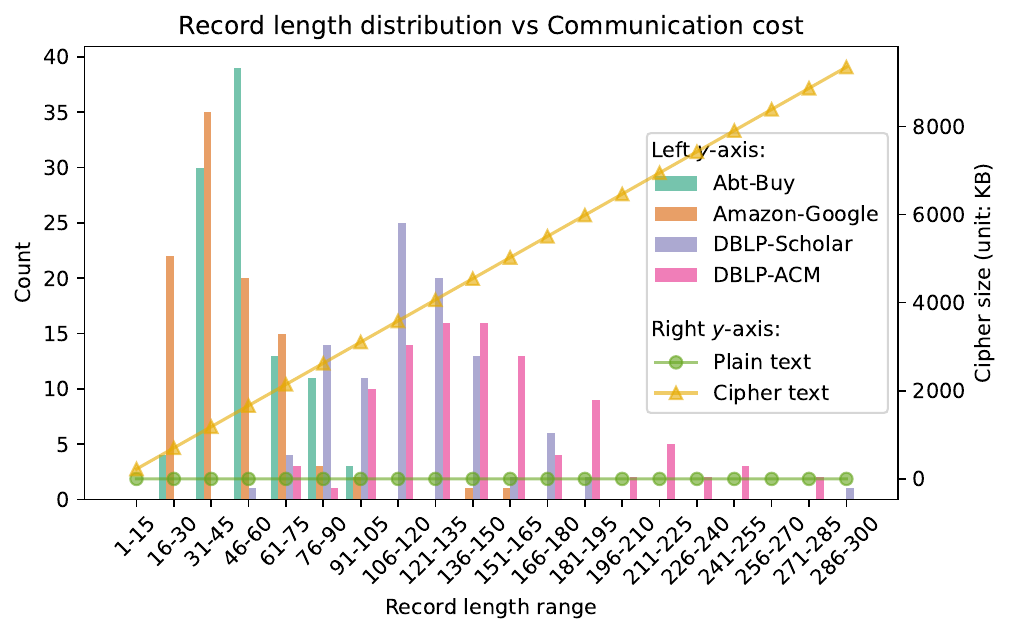}
    \caption{Communication cost in the context of the record length range. The left y-axis indicates the number of such records, and the right y-axis indicates the size of the corresponding plaintext and ciphertext.}
    \label{fig:exp-rec_len-vs-cipher}
\end{subfigure}
\hfill
\begin{subfigure}[t]{.48\textwidth}
  \centering
  \includegraphics[width=1\linewidth]{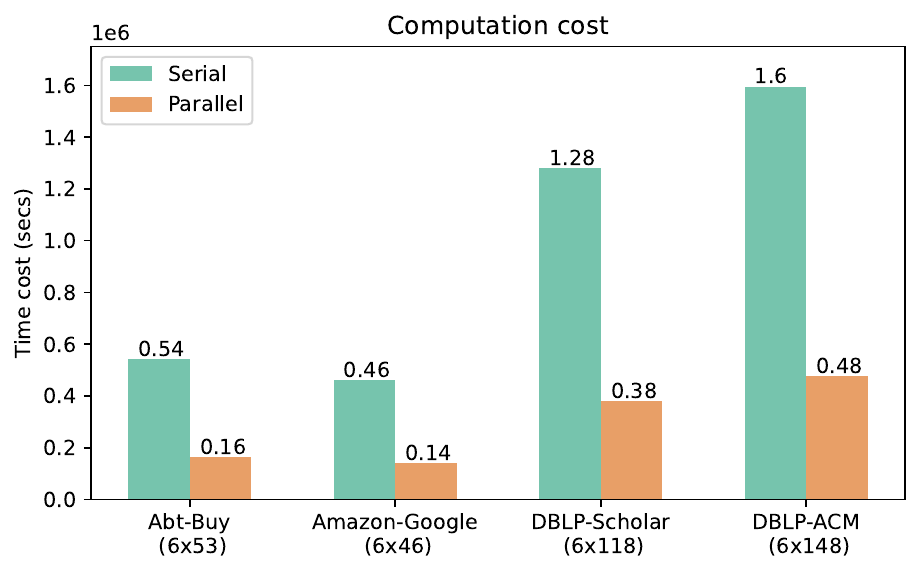}
  \caption{Computation cost in serial and parallel mode. The x-axis represents the length of the two records as $\{\texttt{the average token length} \times \texttt{the average record length}\}$ in each dataset.}
    \label{fig:exp-he-computation-cost}
\end{subfigure}
\end{figure}



Implementing blind annotation based on HE encryption scheme introduces various overheads. We hereby analyze them from the following two perspectives.

\noindent\textbf{Communication cost} We compare the size difference between plaintext and ciphertext with different lengths of input tokens among four datasets. In Figure~\ref{fig:exp-rec_len-vs-cipher}, the x-axis indicates the distribution of record length with a step of 15, the left y-axis indicates the number of records, and the right y-axis indicates the average size of serialized record data in a step in KB. It is clear to find out from the left y-axis that the distribution of the record length is in the range of 1 to 300, and the most common lengths are from 16 to 60 for Abt-Buy and Amazon-Google, and 91-160 for DBLP-Scholar and DBLP-ACM. This leads to the average record length for each dataset is 53, 46, 118, and 148. 
From the right y-axis, the increment of plaintext size, as a baseline, is nuance since each ASCII character only takes 1 byte (0.001KB) and 300 is only 0.3KB. As for the ciphertext, the trend of ciphertext size expands linearly, with an average consumption of 32KB per ASCII character after encryption and serialization. Although the size increment after encryption is significant, the total size of the data is still practical, and such communication is a one-time operation. For example, if the annotation is for 50 records, the average length is 100, then the total size of the encrypted data is 156.25MB.

\noindent\textbf{Computation cost} We evaluate the computation cost on 2500 comparisons (50 records from each party) with the average length of the token size (1st argument, see \Cref{sec:feature-length}) and the average record length (2nd argument) used in \texttt{is\_in} function for each dataset: 6x53, 6x46, 6x118 and 6x148. We also compare the cost with ciphertext in serial and parallel. As is shown in \Cref{fig:exp-he-computation-cost}, the incurred computation cost for ciphertext is about 5.4e5 seconds for 6x53, 4.6e5 seconds for 6x46, 1.28e6 for 6x118 and 1.6e6 seconds for 6x148 in serial mode. The parallel mode executes byte-level equal comparison with independent gate operations in parallel so that the time cost is significantly improved more than 3 times. 

Two optimizations we consider doing in the future are: 1) The records from both parties are encrypted due to the limitation of BinFHE. For some other HE schemes (e.g., BGV/BFV) that allow the HE operations between a plaintext and a ciphertext, only the record from the other party needs to be encrypted so that more computation and communication costs can be further saved.
2) The parallelization is only optimized for the byte-level comparison. Since the 2500 pairwise record comparisons are mutually independent and can also be run simultaneously, the fully paralleled program should cost notably less time.

\section{Conclusion and Future Work}

In this work, we propose a blind annotation protocol based on homomorphic encryption for labeling ground truth that is tailored for privacy-preserving entity resolution. Unlike revamping the traditional annotation methods with de-identified records, this protocol explores the possibility of annotating without revealing any record in plaintext to other parties. The domain-specific language lowers the bar of the implementation in the real world and the comprehensive experiment results show the rationality of blind annotation.
It can be anticipated that scaling efficiently to more complex datasets and real-world scenarios will become increasingly computationally feasible as HE implementations become more efficient.

In the future, because of the extensibility of DSL, more qualified homomorphic evaluation functions could be invented and integrated. Additionally, supporting the multi-attribute dataset is a practical enhancement that increases information utilization and improves linkage accuracy.

\bibliography{citations}
\bibliographystyle{tmlr}

\appendix
\appendix

\section{Preliminaries}
\subsection{Homomorphic Encryption}
\label{sec:more-he}
Homomorphic encryptions are widely used for privacy-preserving computation tasks. For example, users outsourcing their data to a cloud computing platform would have privacy concerns; with homomorphic encryption, the platform is able to perform necessary computations on the encrypted data directly~\cite{naehrig2011can}.

Homomorphic encryption schemes are normally categorized into three types according to the survey~\cite{acar2018survey}: (1) \textit{Partially Homomorphic Encryption} (PHE) allows only one type of operation with an unlimited number of times (i.e., no bound on the number of usages), e.g., Pailier~\cite{paillier1999public}. (2) \textit{Somewhat Homomorphic Encryption} (SWHE) allows some types of operations a limited number of times, e.g., Boneh-Goh-Nissim (BGN)~\cite{boneh2005evaluating}. (3) \textit{Fully Homomorphic Encryption} (FHE) allows an unlimited number of operations for an unlimited number of times. Since Gentry realized the first FHE scheme~\cite{gentry2009fully} based on the lattice, a lot of follow-up FHE schemes have been proposed to address the issue it has and optimized to be practical in real-world applications. Some schemes including Brakerski-Gentry-Vaikuntanathan (BGV)~\cite{cryptoeprint:2011:277}, Brakerski–Fan–Vercauteren (BFV)~\cite{fan2012somewhat,brakerski2012fully} and Cheon-Kim-Kim-Song (CKKS)~\cite{heforarithmeticofapproximatenumbers} support arithmetic operations while some others~\cite{tseng2017homomorphic} including FHEW~\cite{10.1007/978-3-662-46800-5_24} is capable for logical operations. Recently, works are extending the pure arithmetic schemes to run logical comparators without losing advantages such as Single Instruction Multiple Data (SIMD)~\cite{iliashenko2021faster}.

\section{Blind Annotation Protocol}
\subsection{Feature Functions}
\subsubsection{Domain-specific Language}
\label{sec:dsl-syntax}


Some of the features the DSL equips:

\begin{itemize}
    \item The primitive data types are \texttt{string} which is wrapped in double quotes and \texttt{number}.
    \item Three logic operators are supported. Or operator: \texttt{a | b} returns true if either \texttt{a} or \texttt{b} is true. Otherwise, false. And operator: \texttt{a \& b} returns true if both \texttt{a} and \texttt{b} are true. Otherwise, false. Not operator: \texttt{!a} returns true if \texttt{a} is false, or returns false if \texttt{a} is true.
    \item Variables can be defined by \texttt{\$v=\{exp\}} where \$ is the variable indicator, \texttt{v} is the variable name and \texttt{\{exp\}} is a valid expression.
    \item A preset variable \texttt{\$r} is act as the target record for comparison and should be used in the annotation as the argument of the question $Q$.
    \item At least one return statement \texttt{ret} is required. The argument of the return should be in Boolean. If multiple is provided, the code terminates when the first return statement is found.
    \item Round bracket pair \texttt{()} is for prioritizing the execution of specified expressions. 
    \item Comment starts with \texttt{\#}, e.g., \texttt{\# this is a comment}. 
    \item Whitespaces and empty lines are ignored.
\end{itemize}

\subsubsection{Homomorphic Functions}
\label{sec:homomophic-functions}

Besides the challenges mentioned in \Cref{sec:functions} for constructing pre-defined functions, we still need to tackle one more issue that these functions rely on. As \cite{yao2021amppere} pointed out, the control flow including choice (e.g. if) and loop (e.g. for, while) based flows require conditional expressions, which are in Boolean, for execution. These encrypted Boolean values cannot be used in conditional operations unless they are decrypted. However, the decryption of the sensitive encrypted value causes the exposing of the execution path which is not allowed in our privacy-preserving settings. The solution is to import \textbf{ternary operation} that converts to logic operation to be arithmetic operation, hence bypassing the decryption of the encrypted Boolean condition. The ternary operator is defined as \texttt{cond: a ? b} where \texttt{cond} is the encrypted Boolean condition and \texttt{a} or \texttt{b} are encrypted integer values. To return \texttt{a} when \texttt{cond} is true or to return \texttt{b} when \texttt{cond} is false without decrypting \texttt{cond}, we exploit oblivious-style ternary operators~\cite{ohrimenko2016oblivious,constable2018liboblivious} and implement \texttt{choose($\enc{cond},\enc{a},\enc{b}$)} with the arithmetic operators as $\enc{cond} * (\enc{a} - \enc{b}) + \enc{b}$ where $\enc{cond}$, $\enc{a}$ and $\enc{b}$ are all ciphertexts, and the return of \texttt{choose}, which is either $\enc{a}$ or $\enc{b}$, is also a ciphertext.

    


With the ternary operator, we can define more sophisticated functions. \Cref{alg:functions} lists some commonly-used functions and the implementation details. All these functions are evaluated and returned in ciphertexts. \texttt{lower($\enc{s}$)} (line 1-6) and \texttt{upper($\enc{s}$)} (line 7-12) convert the ASCII character to all lower-case or upper-case for standardization. These are not by default applied because, in some circumstances, letter case is an important signal for identifying record similarity. Both methods take in the encrypted string \texttt{s} and loop the characters in it one by one. For each character, it determines if it falls in a certain range of the ASCII table, and uses that as the condition to choose whether to modify the value or not. If the homomorphic scheme supports SIMD operation, calculating \texttt{cond} and \texttt{choose} can be run as a batch in one operation so the for-loop is saved. We also provide \texttt{is\_in($\enc{a},\enc{b}$)} function (line 13-22) which detects if $\enc{a}$ is a sub-string of $\enc{b}$ by scanning the existence of $\enc{a}$ from $\enc{b}$'s left to right. Using this $O(n^2)$ naive method without the early exit, even if a sub-string match exists, is not ideal but unavoidable because of preventing the disclosure of the execution path for privacy, as stated above.

\begin{algorithm}[t]
\small
\caption{Commonly-used Functions}
\label{alg:functions}
    \SetKwFunction{FEnc}{enc}
    \SetKwFunction{FDec}{dec}
    \SetKwFunction{FRand}{rand}
    \SetKwFunction{FLen}{len}
    \SetKwFunction{FChoose}{choose}
    
    \SetKwFunction{FLower}{lower}
    \SetKwFunction{FUpper}{upper}
    \SetKwFunction{FIsIn}{is\_in}
    
    \SetKwInput{KwInput}{Input \& return}
    \KwInput{The input $\enc{s}$, $\enc{a}$, $\enc{b}$ are all homomophically encrypted. The return of all functions is also in ciphertext. $len$ returns the encrypted string length in plaintext.}
    
    \SetKwProg{Fn}{Function}{:}{}
    \Fn{\FLower{$\enc{s}$: string}}{
        \For{$i \gets 1$ \KwTo $len(\enc{s})$}{
            $\enc{c} \gets \enc{s[i]}$\;
            $\enc{cond} \gets (\enc{c} > \mathtt{0x40})\ \wedge\ (\enc{c} < \mathtt{0x5b})$\;
            $\enc{s[i]} \leftarrow \FChoose(\enc{cond}, \enc{c} + \mathtt{0x20}, \enc{c})$\;
        }
        \KwRet $\enc{s}$;\
    }
    \BlankLine
    
    \Fn{\FUpper{$\enc{s}$: string}}{
        \For{$i \gets 1$ \KwTo $len(\enc{s})$}{
            $\enc{c} \gets \enc{s[i]}$\;
            $\enc{cond} \gets (\enc{c} > \mathtt{0x60})\ \wedge\ (\enc{c} < \mathtt{0x7b})$\;
            $\enc{s[i]} \leftarrow \FChoose(\enc{cond}, \enc{c} - \mathtt{0x20}, \enc{c})$\;
        }
        \KwRet $\enc{s}$;\
    }

    
    \SetKwProg{Fn}{Function}{:}{}
    \Fn{\FIsIn{$\enc{a}$: string, $\enc{b}$: string}}{
        $l_a \gets len(\enc{a})$\;
        $l_b \gets len(\enc{b})$\;
        $\enc{res} \gets \enc{\mathtt{False}}$\;

        \For{$j \gets 1$ \KwTo $l_b - l_a + 1$}{
            $\enc{r} \gets \enc{\mathtt{True}}$\;
            \For{$i \gets 1$ \KwTo $l_a$}{
                $\enc{r} \gets \enc{r}\ \wedge\ (\enc{a[i]} = \enc{b[j+i]})$\;
            }
            $\enc{res} \gets \enc{res}\ \vee\ \enc{r}$\;
        }
        \KwRet $\enc{res}$\;
    }
\end{algorithm}

\subsection{Security Analysis}
\label{sec:security}


\subsubsection{Security Definitions}
We model $\mathcal{A}$ as a probabilistic polynomial-time machine and the parties as interactive Turing machines. 

\begin{definition}[Adversary] 
An honest-but-curious adversary $\mathcal{A}$ is capable of corrupting a subset of parties in the system, both not the server and one of the data owners at the same time. A compromised party will divert to $\mathcal{A}$ all the received messages and act as $\mathcal{A}$ requests. 
\end{definition}

\noindent\textbf{Ideal World.} Our ideal-world functionality $\mathcal{F}$ interacts with annotation parties as follows: 

\begin{enumerate}
    \item Each data owner sends $\mathcal{F}$ its plaintext data $r \in D$ and plaintext feature questions $Q$. $\mathcal{F}$ processes and checks Boolean result $w$ as $Q_i(r_j) = Q_j(r_i), \forall Q \forall r$.
    \item If $w$ is true, then add the records from both sides to the ground truth.
    \item If $w$ is false, repeat Step (1) until an agreement is reached or it exceeds a set trial threshold.
\end{enumerate}

 \noindent\textbf{Real World.} In the real world, $\mathcal{F}$ is replaced and realized by our protocol described in the previous parts of this section. 
 
\begin{definition}[Security]
\label{def:security}
A blind annotation protocol is simulation secure if for
every adversary $\mathcal{A}$ in the real world, there exists a security simulator $\mathcal{S}$ in the ideal world that also corrupts the same set of parties and produces an output identically distributed to $\mathcal{A}$'s output in the real world.
\end{definition}

\subsubsection{Security Simulation} 
We describe a security simulator $\mathcal{S}$ that simulates the view of the $\mathcal{A}$ in the real-world execution of our protocol. Our security \Cref{def:security} and $\mathcal{S}$ ensure both confidentiality and correctness. $\mathcal{S}$  receives from  $\mathcal{F}$, $\mathcal{F}(C, x)$, where $C$ is computing circuits. $\mathcal{S}$ sends $\mathcal{F}(C)$ to $\mathcal{S}$ and obtains fake Homomoprhic Encryption circuits $HE_{fake}$. $\mathcal{S}_{HE}$ generates a random string $o_{fake}$ of the same length as output. $\mathcal{S}$ sends $(HE_{fake}, o_{fake})$ to $\mathcal{A}$. As HE circuits distribution is independent, $HE_{fake}$ is computationally indistinguishable from the real HE circuits $HE$ in the real execution. The random output $o_{fake}$ in ideal execution is indistinguishable from $o$ in the real execution. In the ideal world, $\mathcal{S}$ creates fake circuits $HE_{fake}$ and does not use $x$ for computing. Otherwise, $\mathcal{A}$ could use $x$ to evaluate the circuit, which would allow $\mathcal{A}$ to distinguish between real and ideal executions.


\section{Implementation}

\subsection{User Interface}
\label{sec:simulator}

\begin{figure}[!hbt]
    \centering
    \includegraphics[width=1.0\linewidth]{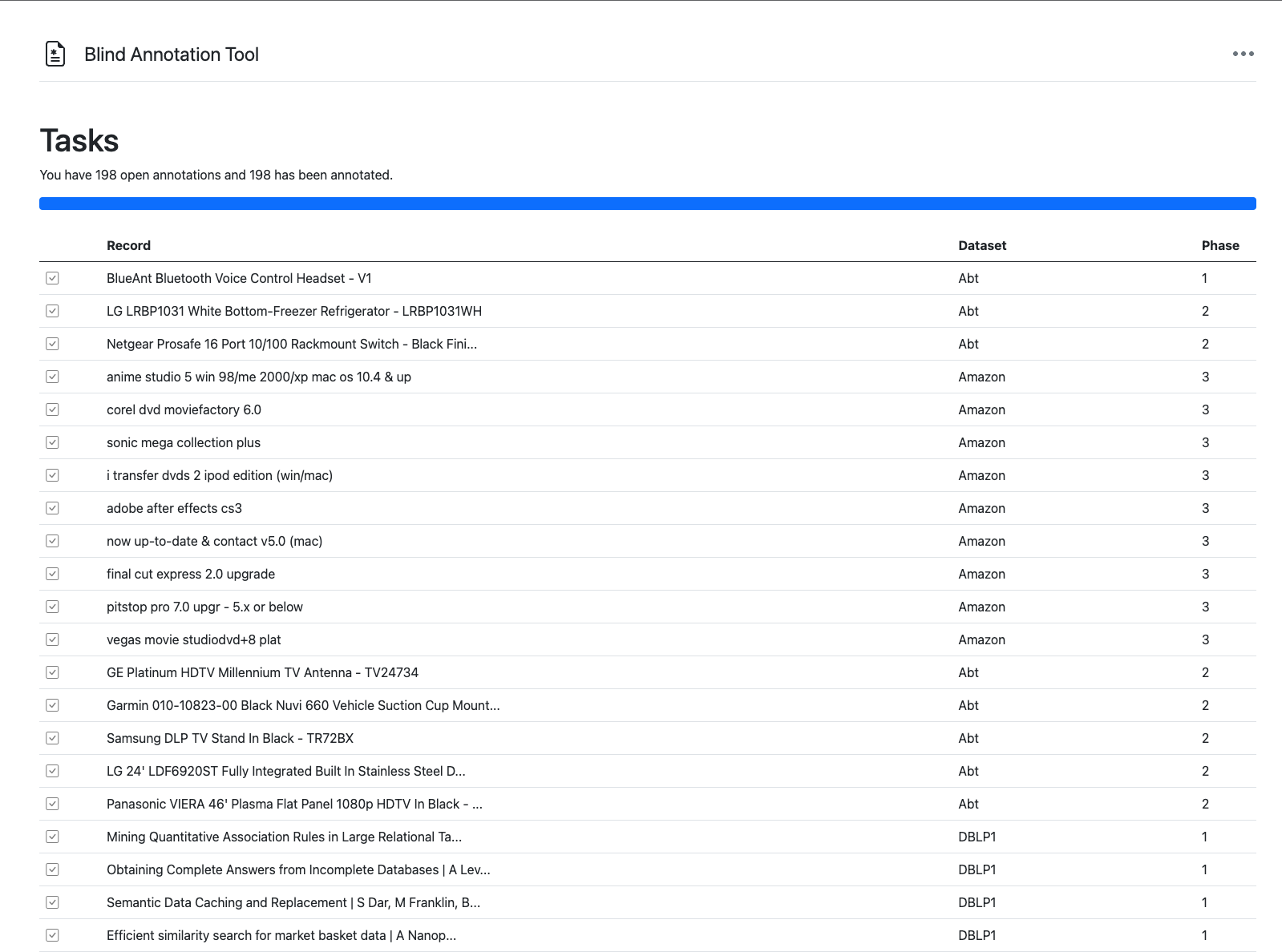}
    \caption{The main page of the web-based GUI}
    \label{fig:anno-tool-1}
\end{figure}

\Cref{fig:anno-tool-1} shows the main page of the web-based GUI which contains the current progress on the top of the page and the records that need to be annotated in the table. For each row of the table, it has an indicator of if the record is annotated or not, a brief of the record content, dataset name, and the current round of annotation for the record.

\begin{figure}[!hbt]
    \centering
    \includegraphics[width=1.0\linewidth]{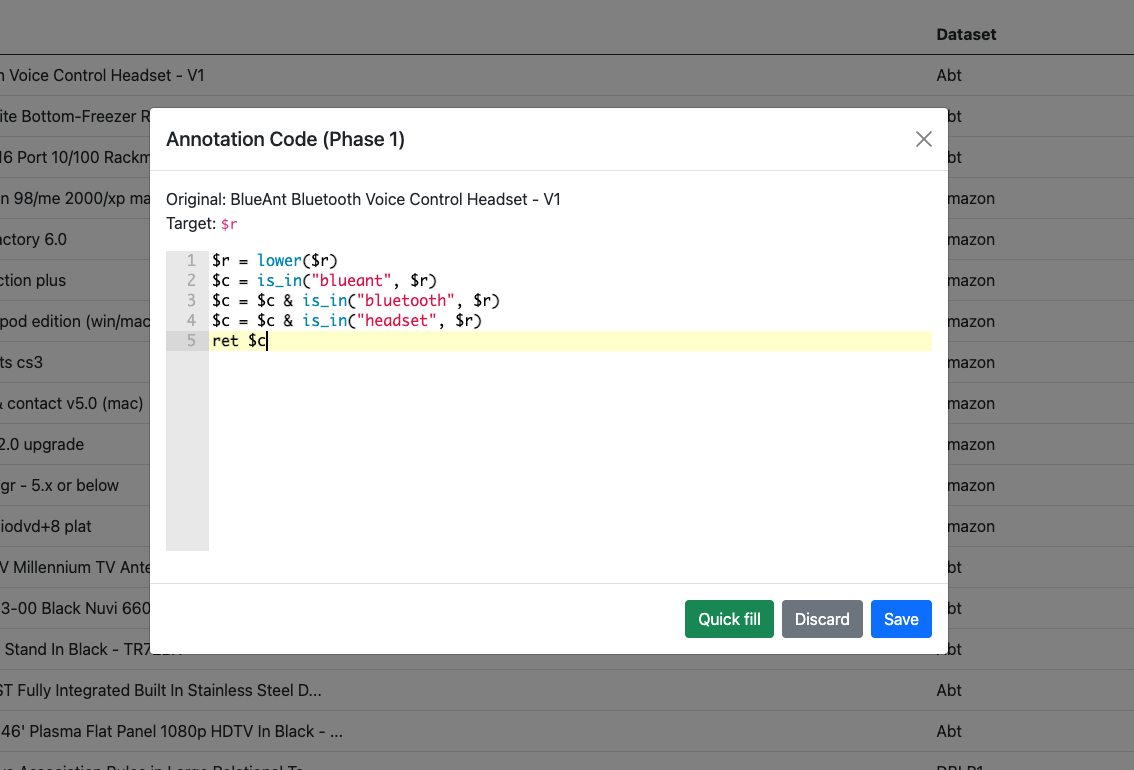}
    \caption{The first round record annotation}
    \label{fig:anno-tool-2}
\end{figure}

When clicking on the record, an annotation window pops up (\Cref{fig:anno-tool-2}): It shows the record content from the dataset that belongs to domain oracles' side, a record placeholder \texttt{\$r} which represents the record from the other data owner side, an annotation editor with DSL code highlighting, and three buttons to quickly fill the editor with an auto-generation heuristic, discard the annotation, and save the annotation respectively. The ``save'' operation also runs the syntax check through the DSL parser to make sure the input is syntactically valid.

\begin{figure}[!hbt]
    \centering
    \includegraphics[width=1.0\linewidth]{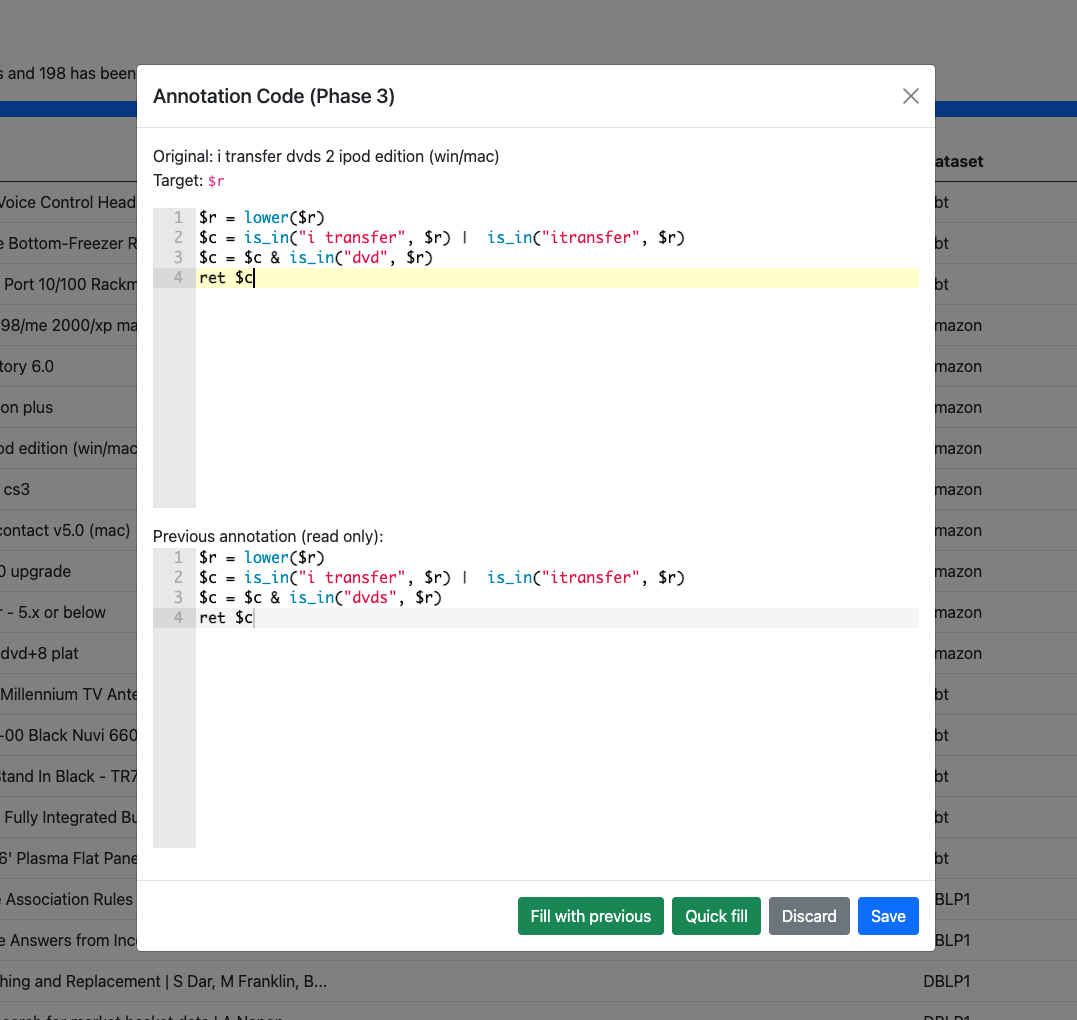}
    \caption{The follow-up round record annotation}
    \label{fig:anno-tool-3}
\end{figure}

When annotating records after the first round, the annotation popup window also shows the previous annotation and allows domain oracles to quickly fill the editor with the previous content (\Cref{fig:anno-tool-3}).

\subsection{Details of the HE implementation}

Each character in a record is ASCII encoded (1 byte), and the Unicode character is converted to the corresponding ASCII character. Therefore, a string is represented by a two-dimensional matrix, where each element in the first dimension represents a character, and each dimension in the second dimension represents 1 bit. For example, a record "Canon" is represented by a 5x8 matrix where the first row is \texttt{01000011} as the binary representation for "C".

Naturally, the encryption is on the bit level since the BinFHE module in OpenFHE defines the operations for bits. Hence each item in the matrix is homomorphically encrypted. The logical operators we employed for implementations from BinFHE are AND ($\wedge$), OR ($\vee$), and XOR ($\oplus$). The detail of the implementation is demonstrated in \Cref{alg:concrete-functions}. Specifically, \texttt{byte\_equal} is the helper function for character-level comparison, which compares each corresponding bit from two encrypted characters $\enc{a}$ and $\enc{b}$. \texttt{is\_in} is the serial version of the function for comparing if the encrypted string $\enc{a}$ is a sub-string of the encrypted string $\enc{b}$. The parallel version utilizes OpenMP and executes character comparison using \texttt{byte\_equal} in parallel.

\begin{algorithm}[t]
\small
\caption{Functions using BinFHE scheme}
\label{alg:concrete-functions}
    \SetKwFunction{FLen}{len}
    \SetKwFunction{FChoose}{choose}
    
    \SetKwFunction{FByteEqual}{byte\_equal}
    \SetKwFunction{FIsIn}{is\_in}
    \SetKwFunction{FIsInPara}{is\_in\_parallel}

    \SetKwProg{Fn}{Function}{:}{}
    \Fn{\FByteEqual{$\enc{a}$: bit[8], $\enc{b}$: bit[8]}}{
        $\enc{res} \gets \enc{\mathtt{True}}$\;

        \For{$j \gets 1$ \KwTo $8$}{
            $\enc{res} = \enc{res} \wedge (\enc{a[i]} \oplus \enc{b[i]})$\;
        }
        \KwRet $\enc{res}$\;
    }
    
    \SetKwProg{Fn}{Function}{:}{}
    \Fn{\FIsIn{$\enc{a}$: [bit[8],$\cdots$], $\enc{b}$: [bit[8],$\cdots$]}}{
        $l_a \gets len(\enc{a})$\;
        $l_b \gets len(\enc{b})$\;
        $\enc{res} \gets \enc{\mathtt{False}}$\;

        \For{$j \gets 1$ \KwTo $l_b - l_a + 1$}{
            $\enc{r} \gets \enc{\mathtt{True}}$\;
            
            \For{$i \gets 1$ \KwTo $l_a$}{
                $\enc{r} \gets \enc{r}\ \wedge\ \FByteEqual(\enc{a[i]}, \enc{b[j+i]})$\;
            }
            $\enc{res} \gets \enc{res}\ \vee\ \enc{r}$\;
        }
        \KwRet $\enc{res}$\;
    }

\end{algorithm}

\section{Experiments}

As the core function for feature extractions in annotation, it is essential to understand the first argument, that is, the input token, in the \texttt{is\_in} function.

\subsection{Feature Length}
\label{sec:feature-length}

\begin{figure}[!hbt]
    \centering
    \includegraphics[width=0.5\linewidth]{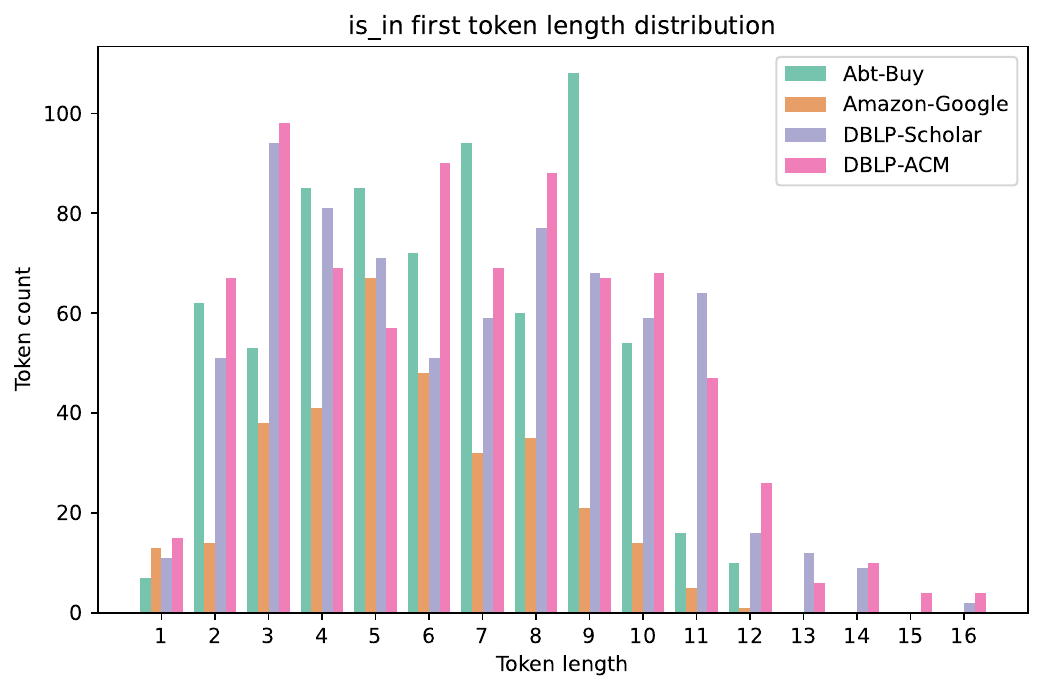}
    \caption{Token length distribution}
    \label{fig:exp-is_in-token-dis}
\end{figure}

The length distribution of the tokens is demonstrated in \Cref{fig:exp-is_in-token-dis}. The length ranges from 1 to 16, and most of the tokens have lengths of 2 to 9. The average length of the token is around 6 (6.65, 6.28, 5.60, 6.61 for Abt-Buy, Amazon-Google, DBLP-Scholar and DBLP-ACM, respectively).

\subsection{Feature Importance}

\begin{figure*}[!hbt]
    \centering
    \includegraphics[width=1.0\linewidth]{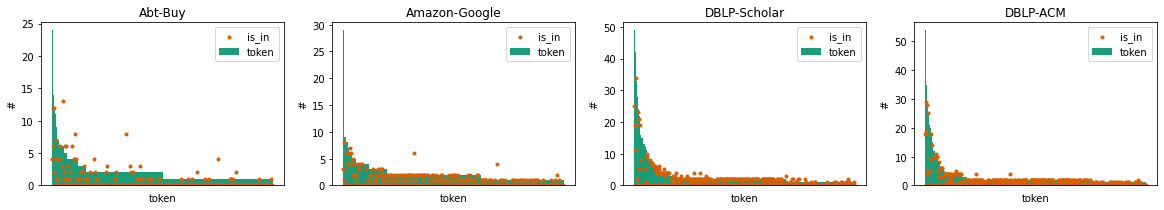}
    \caption{\#tokens appears in the dataset and \#\texttt{is\_in} uses to extract the corresponding tokens from all annotators. The modified tokens that are not in the original token list are discarded. Additionally, token \texttt{-} is removed from Abt-Buy and Amazon-Google for better visualization.}
    \label{fig:exp-is_in-vs-token}
\end{figure*}

We are curious about if the important features are more likely to be extracted by the annotators. We tokenize the record based on white space without any additional processing steps. Meanwhile, we count the number of times that function \texttt{is\_in} is being utilized to extract the corresponding tokens. Note that annotators could modify the tokens accordingly, e.g., ``photoshop'' to be ``ps'', and ``international'' to be ``i18n'', we only use the original token to construct the x-axis, thus none of these non-original tokens are counted if they do not appear in the original token list.

The distribution of tokens and \texttt{is\_in} usage is summarized in \Cref{fig:exp-is_in-vs-token}. The tokens are present in the long-tail form, whereas the dense invocations of \texttt{is\_in} function concentrate on its ``tail''. This observation shows the common tokens contain relatively less information than the rare ones so that the feature extraction depends more on the latter. Some common tokens also attract a fair amount of tokens, these are usually the common but special tokens for records, for instance, the brand name of a product. Some \texttt{is\_in} functions are called a significant amount of times more than the token itself, they derive from the modification of the original token.

\section{Case Study}
\label{sec:case-study}

We selected two cases from two different domains. For both of them, two parties hold different opinions at first but establish an agreement finally. 

The first case in \Cref{fig:case-1} is product records from Amazon-Google dataset. On the Amazon dataset side, the given record content is short. The two attempts of the annotation are identical since not too much useful information can be used. On the Google dataset side, besides the brand and product name, it also has the category ``music production software''. The first annotation has relatively strict rules, whereas the second annotation only tests the brand and the product name.

\begin{figure*}[!hbt!]
\begin{tcolorbox}[colframe=black!70!white,colback=black!10!white]
\begin{multicols}{2}
\textbf{Dataset}: Amazon \\
\textbf{Record}: cakewalk sonar 6 studio \\
\textbf{Annotation 1}:
\begin{lstlisting}
$r = lower($r)
$c = is_in("cakewalk", $r)
$c = $c & is_in("sonar", $r)
ret $c
\end{lstlisting}
\textbf{Annotation 2}:
\begin{lstlisting}
$r = lower($r)
$c = is_in("cakewalk", $r)
$c = $c & is_in("sonar", $r)
ret $c
\end{lstlisting}

\columnbreak

\textbf{Dataset}: Google \\
\textbf{Record}: cakewalk sonar 6 studio edition software music production software \\
\textbf{Annotation 1}: 
\begin{lstlisting}
$r = lower($r)
$c = is_in("cakewalk", $r)
$c = $c & is_in("sonar", $r)
$c = $c & is_in("6", $r)
$c = $c & is_in("studio", $r)
$c = $c & is_in("music", $r)
ret $c
\end{lstlisting}

\textbf{Annotation 2}: 
\begin{lstlisting}
$r = lower($r)
$c = is_in("cakewalk", $r)
$c = $c & is_in("sonar", $r)
ret $c
\end{lstlisting}
\end{multicols}
\end{tcolorbox}
\caption{A case in Amazon-Google}
\label{fig:case-1}
\end{figure*}

The second case is from DBLP-ACM dataset and shown in \Cref{fig:case-2}. The record content from ACM mentioned ``database'', but in DBLP, it is represented as ``DBMS''. One possible solution is that annotators could expand ``DBMS'' to be ``database'' with their domain knowledge. However, the record content is about the paper title and it usually remains consistent across different websites, so the annotators do not modify it on the token level. Wisely, the annotation from the DBLP's side splits the paper title into two parts, the first part before ``:'' is as the first attempt, and the rest is as the second attempt.

\begin{figure*}[!hbt!]
\begin{tcolorbox}[colframe=black!70!white,colback=black!10!white]
\begin{multicols}{2}
\textbf{Dataset}: DBLP \\
\textbf{Record}: ``Honey, I Shrunk the DBMS'': Footprint, Mobility, and Beyond (Panel) | Praveen Seshadri | SIGMOD Conference | 1999 \\
\textbf{Annotation 1}:
\begin{lstlisting}
$r = lower($r)
$c = is_in("honey", $r)
$c = $c & is_in("i", $r)
$c = $c & is_in("shrunk", $r)
$c = $c & is_in("the", $r)
$c = $c & is_in("dbms", $r)
ret $c
\end{lstlisting}
\textbf{Annotation 2}:
\begin{lstlisting}
$r = lower($r)
$c = is_in("footprint", $r)
$c = $c & is_in("mobility", $r)
$c = $c & is_in("and", $r)
$c = $c & is_in("beyond", $r)
ret $c
\end{lstlisting}

\columnbreak

\textbf{Dataset}: ACM \\
\textbf{Record}: Honey, I shrunk the database: footprint, mobility, and beyond | Praveen Seshadri | International Conference on Management of Data | 1999 \\
\textbf{Annotation 1}: 
\begin{lstlisting}
$r = lower($r)
$c = is_in("honey", $r)
$c = $c & is_in("shrunk", $r)
$c = $c & is_in("the", $r)
$c = $c & is_in("footprint,", $r)
$c = $c & is_in("mobility,", $r)
$c = $c & is_in("and", $r)
$c = $c & is_in("beyond", $r)
\end{lstlisting}

\textbf{Annotation 2}: 
\begin{lstlisting}
$r = lower($r)
$c = is_in("honey", $r)
$c = $c & is_in("shrunk", $r)
$c = $c & is_in("the", $r)
$c = $c & is_in("footprint", $r)
$c = $c & is_in("mobility,", $r)
$c = $c & is_in("and", $r)
$c = $c & is_in("beyond", $r)
$c = $c & is_in("seshadri", $r)
ret $c
\end{lstlisting}
\end{multicols}
\end{tcolorbox}
\caption{A case in DBLP-ACM}
\label{fig:case-2}
\end{figure*}

\end{document}